\documentclass[%
 reprint,onecolumn,
 amsmath,amssymb,
 aps,
]{revtex4-1}
\usepackage{color}
\usepackage{graphicx}
\usepackage{dcolumn}
\usepackage{bm}
\usepackage{filecontents}

\begin{document}
\preprint{APS/123-QED}

\title{\bf Polarization of a probe laser beam due to nonlinear QED effects}

\author
       {Soroush Shakeri $^{1,2}$\footnote{E-mail: Soroush.Shakeri@ph.iut.ac.ir}, Seyed Zafarollah Kalantari$^1$\footnote{E-mail: zafar@cc.iut.ac.ir}, She-Sheng Xue$^{2,3}$\footnote{E-mail: xue@icra.it}
       \\
       $^1$Department of Physics, Isfahan University of Technology, Isfahan 84156-83111, Iran\\
       $^2$ICRANet Piazzale della Repubblica, 10 -65122, Pescara, Italy\\
       $^3$Physics Department,
University of Rome La Sapienza, P.le Aldo Moro 5, I–00185 Rome, Italy\\ 
       }

\date{\today}

\begin{abstract}
Nonlinear QED interactions  induce different polarization properties on a given probe beam. We consider the polarization effects caused by the photon-photon interaction in laser experiments, when a laser beam propagates through a constant magnetic field or collides with another laser beam. We solve the quantum Boltzmann equation  within the framework of the Euler-Heisenberg Lagrangian for both time-dependent and constant background field to explore the time evolution of the Stokes parameters \emph{Q}, \emph{U} and \emph{V} describing polarization. Assuming an initially linearly polarized probe laser beam, we also calculate the induced ellipticity and rotation of the polarization plane.
\end{abstract}

\pacs{Valid PACS appear here}
\maketitle
\begin{center}
\section{Introduction}
\end{center}
The strong-field  regime of QED is a completely new area of physics with new phenomena such as "nonlinear QED effects"; these features will be tested by ongoing experiments. The QED quantum vacuum made up of virtual pairs, when exposed to  intense light,  effectively behaves  as a birefringent medium \cite{1935NW.....23..246E,1936ZPhy...98..714H,2014PhRvD..90d5025D,1931ZPhy...69..742S}, due to  nonlinear interactions introduced into the linear Maxwell's equation. Recently, there have been many experimental efforts  to detect photon-photon interactions \cite{HIBEF,ELI,2008PhRvD..77c2006Z,2016EPJC...76...24D,2013NJPh...15e3026D,1998ftqp.conf...29D}, the direct evidence of elastic scattering has not been observed yet. These experiments are mainly based on ellipticity induced in an initially linearly polarized probe laser  passing through a  background field, provided by  superconducting magnets \cite{2008PhRvD..77c2006Z,2016EPJC...76...24D,2013NJPh...15e3026D,1998ftqp.conf...29D} or high intensity lasers \cite{HIBEF,ELI,2012RvMP...84.1177D,2015FBS....56..615T} in analogy to a  beam of light passing through a birefringent crystal. 
This means we have two cases for analysis: (1) the laser propagation through a static magnetic field  made by ground magnets and (2) the collision between two laser beams in which one of the laser beam behaves as an electromagnetic background for another one.

In order to observe the QED nonlinearities for weak fields in the laboratory, long interaction length or time is required. In these kinds of experiments \cite{2008PhRvD..77c2006Z,2016EPJC...76...24D,2013NJPh...15e3026D}, there is an interaction of optical laser photons, which are highly linearly polarized, with the magnetic-field strength  of the order of a few tesla. By contrast, upcoming high intensity laser systems will be able  to achieve ultra-high-field strengths in the laboratory \cite{2012RvMP...84.1177D,2009EPJD...55..311G,2008arXiv0809.3348H,Heinzl:2008kv}.  The current laser intensity record (in the optical regime) is given by $2\times 10^{22}{W}/cm^{2}$ \cite{2008OExpr..16.2109Y} and future facilities envisage even higher intensities of the order of $ 10^{24}-10^{25}{W}/{cm^{2}}$ \cite{ELI}. Because of this drastic enhancement of field strengths, it is possible to probe distances of about electron Compton wave length where the  quantum effects play an important role \cite{2008arXiv0809.3348H}. There is a critical electric field $E_{cr}=1.3\times10^{16}{V}/{cm}$, in which an electron gains an energy $m_{e}$ upon traveling a distance equal to its Compton wavelength \cite{2009OptCo.282.1879H,2008arXiv0809.3348H,Heinzl:2008kv}. In such a field it becomes energetically favourable to produce real electron-positron pairs from the vacuum \cite{1931ZPhy...69..742S,1936ZPhy...98..714H,1951PhRv...82..664S,2010PhR...487....1R}, and  correspondingly a critical magnetic field $B_{cr}=4.4\times 10^{13}$ G. In \cite{2006OptCo.267..318H,2006PhRvL..97h3603D,2014PhRvA..89f2111M} there are some proposals  to measure QED nonlinearities, especially in \cite{2006PhRvL..97h3603D}, where by using combined optical and x-ray lasers, due to the birefringence effect, QED nonlinearities increase with background field strength and probe beam frequency. This scenario will be realized with the Helmholtz international
beamline for extreme fields (HIBEF) facility employing the European x-ray free electron laser (XFEL) at DESY \cite{HIBEF}, and will, if successful, give the first experimental verification of vacuum birefringence along with photon-photon scattering. 

However, it is not clear if one gets the same results from the experiments in the static strong-field or time-dependent one. The laser beam consists of oscillating electric and magnetic fields, and the authors  of \cite{2006OptCo.267..318H,2006PhRvL..97h3603D} have proposed  that a  standing wave generated by the superposition of two counter propagating strong and
tightly focused optical laser beams can be used as a target for a given x-ray probe beam. Although in principle it is possible to setup high power short pulse laser standing waves, this would not be easily  achieved in practice. In this paper, in addition to  the static strong field, we consider a time-dependent field configuration which occurs in the cross region of two laser beams the frequencies of which are different. The aims of this paper are (i) presenting a quantum-mechanical description  for the generation of an ellipticity signal in both constant and time-varying background fields; (ii) investigating observational consequence and discrepancy of these two cases; and  (iii) considering the effects of both time-varying electric and magnetic fields on the polarization characteristics of a given probe beam. For these purposes, we use the generalized quantum Boltzmann equation formulated in \cite{2009PhRvD..79f3524A,1996AnPhy.246...49K}, to study the evolution of polarization characteristics for a probe beam propagating through background fields. In this paper we have extended the recent study \cite{2014PhRvA..89f2111M} to provide more details of the different experimental setups in more realistic configurations.
This paper is organized as follows. In Sec. II, we review  prescription of the polarization by Stokes parameters and relating to ellipticity and rotation of polarization plane angles, and the evolution in time obeying  the quantum Boltzmann equation. In Sec. III, we examine the generation of polarization characteristics of the probe beam in the presence of the background fields, taking into account the Euler-Heisenberg effective Lagrangian. In Sec. IV, we obtain the set of equations governing the time evolution of the Stokes parameters.  We analytically and numerically  solve these equations to obtain polarization characteristics of the probe beam in both cases of time-dependent and static background fields. Finally, in the last section after some discussions, we give some concluding remarks.
\newpage

\section{Faraday rotation, Stokes parameters and Boltzmann equation}\label{II}
We consider a  monochromatic electromagnetic wave propagating in the $\hat z$ direction, where the electric and magnetic fields oscillate in the \emph{x}-\emph{y} plane. At a spatial point, the electric-field vector can be written as 
\begin{align}\label{e03}
\vec E=E_{x}\hat x +E_{y}\hat y=(E_{0x}e^{i\phi_{x}}\hat x \ + E_{0y}e^{i\phi_{y}}\hat y)e^{-i\omega t}
\end{align}
Where, $\phi_{x}$ and $\phi_{y}$ are the phases at the initial time, and $E_{0x}$ and $E_{0y}$ are amplitudes in the $\hat x $ and $\hat y$ directions. Polarization properties are normally described in terms of the Stokes parameters: the total intensity \emph{I}, linear polarization \emph{Q} and \emph{U}, and the circular polarization \emph{V}. They are defined as time-averaged quantities \cite{Jackson:1975up,2009PhRvD..79f3524A,1996AnPhy.246...49K,2003PhLB..554....1C}.
\begin{align}\label{e1}
I\equiv\langle E_{x}^{2}\rangle +\langle E_{y}^{2}\rangle \ ,
\end{align}
\begin{align}\label{e2}
Q\equiv\langle E_{x}^{2}\rangle -\langle E_{y}^{2}\rangle \ ,
\end{align}
\begin{align}\label{e3}
U\equiv\langle2 E_{x} E_{y}\cos(\phi_{x}-\phi_{y})\rangle \ ,
\end{align}
\begin{align}\label{e4}
V\equiv\langle2 E_{x} E_{y}\sin(\phi_{x}-\phi_{y})\rangle\ ,
\end{align}
\emph{Q} and \emph{U} depend on the orientation of the coordinate system  on the plane orthogonal to the direction of propagation, instead  \emph{I} and \emph{V} are  independent of this coordinate system \cite{2009PhRvD..79f3524A}.

 Alternatively,  linearly polarized light can  be considered as a superposition of two opposite circular polarized waves,
\begin{align}\label{e04}
\vec E=E_{R}\hat R+E_{L}\hat L=(E_{0R}e^{i\phi_{R}}\hat R \ + E_{0L}e^{i\phi_{L}}\hat L)e^{-i\omega t},
\end{align}
where $\hat R$($\hat L$) stands for unit vectors for the right-hand (left-hand)  polarization. According to this redefinition \emph{Q} and \emph{U} can be expressed as
\begin{align}\label{e05}
Q\equiv\langle2 E_{R} E_{L}\cos(\phi_{R}-\phi_{L})\rangle,
\end{align}
\begin{align}\label{e06}
U\equiv\langle2 E_{R} E_{L}\sin(\phi_{R}-\phi_{L})\rangle.
\end{align}
If a wave propagates through a magnetized plasma, its polarization vectors will rotate. Then the net rotation of the plane of polarization is $\Delta \phi_{FR}=\frac{1}{2}\ (d\phi _{L}-d\phi _{R})=(k_{L}-k_{R})dz$, this phenomenon is called Faraday rotation (FR). This phase shift mixes \emph{Q} and  \emph{U} parameters such as
\begin{align}\label{e07}
\dot Q=-2U\frac{d\Delta\phi_{FR}}{dt},
\end{align}
\begin{align}\label{e08}
\dot U=-2Q\frac{d\Delta\phi_{FR}}{dt}.
\end{align}
On the other hand, when linear polarized light with circular polarized normal modes passes through an optically active sample with a different absorbance  for different components (refer to the imaginary part of a medium's refractive index),  circular polarized light is a  consequence of this absorbance.
For an initially linearly polarized wave with linear components as normal modes like Eq (\ref{e03}), in contrast to  decomposition to the circular states, the difference in phase velocities $\phi_{x}-\phi_{y}$ leads to a mixing between   \emph{U} and  \emph{V} parameters 
\begin{align}\label{e09}
\dot V=2U\frac{d\Delta\phi_{FC}}{dt},
\end{align}
which measures the  angle related to Farady conversion (FC) in a magnetized medium\cite{2003PhLB..554....1C}.

The quantum vacuum  in the background fields effectively behaves as a birefringence medium, introducing  a faraday rotation (circular birefringence) on the light beam passing through it\cite{1998ftqp.conf...29D}. The complex index of refraction is written as $\tilde n=n+i\kappa$ where n is the index of refraction and $\kappa$ is the extinction coefficient. A linear birefringence can be described as the difference between the real refraction indices for the two polarizations $\Delta n=n_{\parallel}-n_{\perp}$, where $n_{\parallel}$ is parallel to optical axis (in our case the magnetic field direction) and $n_{\perp}$ is perpendicular to it. Similarly,  a dichroism can be defined as the difference in extinction coefficient $\Delta \kappa=\kappa_{\parallel}-\kappa_{\perp}$\cite{2008PhRvD..77c2006Z}. During propagation along a path with length L , a birefringence $\Delta n$ and a dichroism $\Delta \kappa$ generate an ellipticity $\epsilon$ and a rotation angle of the major axis of the ellipse $\psi$ respectively, which  are represented by
\begin{align}\label{e09a}
\epsilon=\frac{\pi\Delta n L}{\lambda}\ \sin 2\vartheta,
\end{align}
\begin{align}\label{e010}
\psi=\frac{\pi\Delta\kappa L}{\lambda}\ \sin 2\vartheta,
\end{align}
where $\vartheta$ is the angle between the light polarization vector and the optical axis direction.  The Stokes parameters can be related to the polarization rotation angle ($\psi$)  and ellipticity angle ($\epsilon$) \cite{Collett:2005cm},
\begin{align}\label{e011}
I=I_{0},
\end{align}
\begin{align}\label{e012}
Q=I_{0}\cos2\epsilon\cos2\psi,
\end{align}
\begin{align}\label{e013}
U=I_{0}\cos2\epsilon\sin2\psi,
\end{align}
\begin{align}\label{e014}
 V=I_{0}\sin2\epsilon,
\end{align}
where $I_{0}$, $2\psi$ and $2\epsilon$ are the spherical coordinates of the three-dimensional vector of Cartesian coordinates (\emph{Q}, \emph{U}, \emph{V}). Since the three parameters $I_{0}$, $2\psi$, and $2\epsilon$ determine the four Stokes parameters, there must be a relation between the Stokes
parameters, this relation is $I^{2}=Q^{2}+U^2+V^2$ (only true for the $100\%$ polarized light). For  a general source of
light which is  a superposition of many different waves without any  fixed phase relation between them,  all four Stokes parameters are independent of each others and should be measured separately, in this case $I^{2}\geq Q^{2}+U^2+V^2$. However, for  given sets of  Stokes parameters, one can transform the spherically coordinate Eqs. (\ref{e011})-(\ref{e014}) to obtain the following relations:
\begin{align}\label{e015}
I=I_{0},
\end{align}
\begin{align}\label{e016}
2\psi=\arctan(\frac{U}{Q} ),
\end{align}
\begin{align}\label{e017}
2\epsilon=\arctan(\frac{V}{\sqrt{Q^{2}+U^{2}}} ).
\end{align}
Generally if one has a polarization ellipse that the electric-field end point traces out, $\psi$ represents the angle of the polarization plane and  $\epsilon$ manifests its ellipticity.

The Stokes parameters can be defined in a quantum mechanical description by using the quantum operators and states  in which  the linear basis corresponds to  Stokes parameters \cite{2009PhRvD..79f3524A,1996AnPhy.246...49K}. In a general mixed state, an ensemble of photons can be described by a normalized density matrix $\rho_{ij}\equiv(\left\vert \epsilon_{i}\right\rangle\left\langle \epsilon_{j} \right\vert/\mathrm{tr\rho})$. The density matrix $\rho$ on the polarization state space encodes the intensity and polarization of the photon ensemble; the  expectation value for the Stokes parameters is given by 
\begin{align}\label{e10}
\mathrm{I} \equiv \left\langle \hat I \right\rangle= \mathrm{tr }{\rho\hat I }=\rho_{11}+\rho_{22},
\end{align}
\begin{align}\label{e11}
\mathrm{Q} \equiv \left\langle \hat Q \right\rangle= \mathrm{tr }{\rho\hat Q }=\rho_{11}-\rho_{22},
\end{align}
\begin{align}\label{e12}
\mathrm{U} \equiv \left\langle \hat U \right\rangle= \mathrm{tr }{\rho\hat U }=\rho_{12}+\rho_{21},
\end{align}
\begin{align}\label{e13}
\mathrm{V} \equiv \left\langle \hat V \right\rangle= \mathrm{tr }{\rho\hat V }=i(\rho_{12}-\rho_{21}).
\end{align}
These relations show that the density matrix for a system of photons contains the same information as the four Stokes parameters. The explicit expression of the density matrix in the linear polarization basis in terms of Stokes parameters is given as follows
\begin{align}\label{e14}
\rho =\frac{1}{2}\begin{pmatrix}I+Q & U-iV \\
U+iV & I-Q \\
\end{pmatrix}.
\end{align}
In fact, one can evaluate the time evolution of  the polarization vectors by considering the time evolution of the density matrix. The density operator $\hat \rho$ for an ensemble of free photons is given by 
\begin{align}\label{e15}
\hat \rho =\int \frac{d^{3}p}{(2\pi)^{3}}\rho_{ij}(p) \hat a_{i}^{\dag}(p)\hat a_{j}(p),
\end{align}
where the density matrix elements $\rho_{ij}$ are related to the number operators $ \mathcal{\hat {D}}_{ij}(\mathbf{k})=\hat a_{i}^{\dag}(\mathbf{k})\hat a_{j}(\mathbf{k})$  as follows

\begin{align}\label{e16}
\langle \mathcal{\hat {D}}_{ij}(\mathbf{k})\rangle=(2\pi)^{3}2k^{0}\delta^{(3)}(0)\rho_{ij}(\mathbf{k}).
\end{align}
The time evolution of the  photon number operator  can be given as
\begin{align}\label{e17}
\frac{d}{dt}\mathcal{\hat {D}}_{ij}=i[\hat H,\mathcal{\hat {D}}_{ij}],
\end{align}
\begin{align}\label{e18}
\langle\frac{d}{dt}\mathcal{\hat {D}}_{ij}\rangle(t)\simeq i\langle[\mathcal{\hat H}_{int} ,\mathcal{\hat {D}}_{ij}]\rangle-\int_{0}^{t}dt^{'}\langle[\mathcal{\hat H}_{int} (t-t^{'}),[\mathcal{\hat H}_{int} (t^{'}),\mathcal{\hat {D}}_{ij}]]\rangle,
\end{align}
where $\hat H$ is the total Hamiltonian and $\mathcal{\hat H}_{int}$ is the interacting Hamiltonian. Equation (\ref{e18}) can be expressed in terms of the density matrix \cite{1996AnPhy.246...49K}
\begin{align}\label{e19}
(2\pi)^{3}2k^{0}\delta^{(3)}(0)\frac{d}{dt}\rho_{ij}(\mathbf{k})=i\langle[\mathcal{\hat H}_{int} (t),\mathcal{\hat {D}}_{ij}(\mathbf{k})]\rangle-\frac{1}{2}\int^{+\infty}_{-\infty} dt \langle[\mathcal{\hat H}_{int} (t),[\mathcal{\hat H}_{int} (0),\mathcal{\hat {D}}_{ij}(\mathbf{k})]]\rangle.
\end{align}
Equation (\ref{e19}) is called the quantum Boltzmann equation, and  the time evolution of the Stokes parameters is given by this equation; the first term on the right-hand side is referred to as the forward-scattering term while the second one is the 
usual collision term. In the case of linear Maxwellian electrodynamics, the time evolution of the Stokes parameter \emph{V} is always equal to zero, and no  circular polarization occurs. In the following section, we will show that if nonlinear QED interaction in the presence of strong background fields is included the time evolution of the \emph{V} does not vanish leading to non-zero circular polarization. According to Eqs. (\ref{e017}) and (\ref{e016}), the  time dependence of \emph{V}, \emph{U} and \emph{Q} parameters give rise to the  variation of ellipticity $\epsilon$ and a rotation of polarization plane $\psi$ which  are principally measurable quantities in laser experiments.

\section{Euler-Heisenberg Lagrangian and the Generation of circular polarization in the presence of background fields} 
\begin{figure}[htp]
\vspace{-2cm}
\includegraphics[scale=.6]{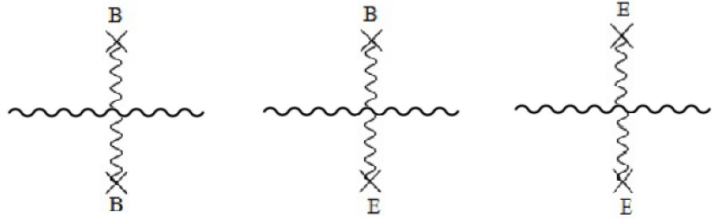}
\vspace{-12cm}
\caption{\scriptsize Photon-photon interactions in the background of  electromagnetic fields, where B represents the classical magnetic field and E represents the classical electric field.  }\label{ff1}
\end{figure}

Here we are going  to consider the nonlinear QED process in the strong background fields  which are   provided by superconducting magnets or by an ultra intense laser. In fact an intense laser field represents a photon coherent state with a large number of photons. In a typical petawatt class laser there are $10^{18}$ photons in a cubic laser wave length;  the correspondence principle tells us this laser beam is very well behaved like classical electromagnetic fields \cite{1951PhRv...82..664S}.
In the following, we adopt the effective Euler-Heisenberg  Lagrangian at the level of one-loop calculation \cite{1936AnP...418..398E,1936ZPhy...98..714H,1998PhRvD..57.2443D,Dunne:2012hp} 
\begin{align}\label{e20}
\mathcal{L}_{eff}=-\frac{1}{4}\mathcal{F}_{\mu\nu}\mathcal{F}^{\mu\nu}+\frac{1}{180}\frac{\alpha^{2}}{m^{4}_{e}}[5(\mathcal{F}_{\mu\nu}\mathcal{F}^{\mu\nu})^{2}-14\mathcal{F}_{\mu\nu}\mathcal{F}^{\nu\lambda}\mathcal{F}_{\lambda\rho}\mathcal{F}^{\rho\mu}],
\end{align}
where the first term is the classical Maxwell Lagrangian, $m_{e}$ is the electron mass and $\alpha$ is the fine-structure constant. In the  Euler-Heisenberg Lagrangian the photon interactions in the presence of a background field can be considered by replacing  $\mathcal{F}_{\mu\nu}\rightarrow f_{\mu\nu}+F_{\mu\nu}$ with the probe quantum field $f_{\mu\nu}$ as a weak perturbation on top of the strong background field $F_{\mu\nu}$, which varies slowly in comparison with $f_{\mu\nu}$.  The quantum field $ f_{\mu\nu}=\partial_{\mu}\hat A_{\nu}-\partial_{\nu}\hat A_{\mu}$ and the
 free photon field $\hat A_{\mu}$  in the Coulomb (radiation) gauge can be written as 
\begin{align}\label{e21}
\hat A_{\mu}(x)=\int\frac{d^{3}\mathbf{k}}{(2\pi)^3 2k^{0}}[\hat a_{r}(k)\epsilon_{r\mu}
(k)e^{-ik.x}+\hat a_{r}^{\dagger}(k)\epsilon_{r\mu}^{\ast}(k)e^{ik.x}],
\end{align}
where $\epsilon_{r\mu}(k)=(0,\vec \epsilon_{r}(k))$, \emph{r}=1,2 shows the photon polarization four-vectors for the two orthogonal transverse polarizations and $k$ (with $k^{0}=\left\vert\mathbf{k}  \right\vert$) stands for the  four-momentum. The  creation and annihilation operators satisfy the canonical commutation relation as
\begin{align}\label{e22}
[\hat a_{i}(k) , \hat a_{j}^{\dagger}(k^{'})]=(2\pi)^{3}2k^{0}\delta_{ij}\delta^{(3)}(\mathbf{k}-\mathbf{k}^{'}).
\end{align}
We compute the evolution of the density matrix and  the Stokes parameters in the effective Euler-Heisenberg (\ref{e20}) framework and in the presence of background fields. We only consider the leading  term (forward-scattering) in the Boltzmann equation (\ref{e19}) as follows
\begin{align}\label{e222}
(2\pi)^{3}2k^{0}\delta^{(3)}(0)\frac{d}{dt}\rho_{ij}(\mathbf{k})=i\langle[\mathcal{\hat H}_{int} (t),\mathcal{\hat {D}}_{ij}(\mathbf{k})]\rangle.
\end{align}
\\
Regarding the canonical commutation relations of the creation and annihilation operators  and their expectation values given in \cite{1996AnPhy.246...49K} it is easy to see that the non-vanishing contribution to the right-hand side of Eq. (\ref{e222})  comes from the photon interaction with two background-field $F_{\mu\nu}$  as shown in Figure (\ref{ff1}). The interaction with three background-fields  has no contribution to the term  $\langle[\mathcal{\hat H}_{int} (t) \mathcal{\hat {D}}_{ij}(\mathbf{k})]\rangle$ of Eq. (\ref{e222}), because  we have ignored  correlations such as $\langle\hat a_{i}(k)\hat a_{j}(k^{'})\rangle$ and $\langle\hat a^{\dagger}_{i}(k)\hat a^{\dagger}_{j}(k^{'})\rangle$.  In fact,  we are assuming that the background field $F_{\mu\nu}$ varies slowly enough in time so that physical two-photon states are neither created nor destroyed by the interaction.   In this case the nonlinear interacting part of the Lagrangian (\ref{e20})  can be found as follows
\begin{align}\label{e23}
\mathcal{L}_{int}=\frac{\alpha^{2}}{90m^{4}_{e}}[5f_{\mu\nu}f^{\mu\nu}F_{\lambda\rho}F^{\lambda\rho}+10F_{\mu\nu}f^{\mu\nu}f_{\lambda\rho}F^{\lambda\rho}-14f
_{\mu\nu}F^{\nu\lambda}f_{\lambda\rho}F^{\rho\mu}-28f_{\mu\nu}f^{\nu\lambda}F_{\lambda\rho}F^{\rho\mu}].
\end{align}

By substituting (\ref{e21}) into (\ref{e23}) and using  the coulomb gauge ($\hat\epsilon\cdot \hat k=0$), one can calculate $\mathcal{H}_{int}=-\int d^{3}x\mathcal{L}_{int}$ as follows

\begin{align}\label{e245}
\mathcal{H}_{int}=\frac{2\alpha^{2}}{90m^{4}_{e}}[\int\frac{d^{3}k}{(2\pi)^{3}(2k^{0})^{2}}\sum_{ss^{'}}\hat a^{\dagger}_{s}(k)\hat a_{s^{'}}(k)(12k^{\mu}F_{\mu\nu}\epsilon_{s^{'}}^{\nu}k^{\lambda}F_{\lambda\rho}\epsilon^{\ast\rho }_{s}+
28k^{\lambda}F_{\lambda\rho}F^{\rho\mu}k_{\mu})].
\end{align}
According to this interaction,  the background field $F_{\mu\nu}$ behaves as an effective coupling between photons. In other words the photons interact with each other via background fields. We now proceed to evaluate $\langle[\mathcal{H}^{B,2}_{int}(t),\mathcal{\hat {D}}_{ij}(\mathbf{k})]\rangle$ by using the contraction relation:
\begin{align}\label{e246}
\langle \hat a_{s^{'}}^{\dagger}(k^{'})\hat a_{s}(k)\rangle=2k^{0}(2\pi)^3\delta^{3}(\mathbf{k}-\mathbf{k}^{'})\rho_{s^{'}s}(\mathbf{k}).
\end{align} 
There is no contribution from the second part of Eq. (\ref{e245}) which contains only the background fields. Consequently the time evolution for the density matrix Eq. (\ref{e222}) can be obtained as follows 
\begin{align}\label{e25}
\frac{d\rho_{ij}}{dt}=\frac{2i\alpha^{2}}{15k^{0}m^{4}_{e}}[\sum_{s} ( k^{\mu}F_{\mu\nu}\epsilon_{s}^{\nu} \ k^{\lambda} F_{\lambda\rho}\epsilon^{\ast\rho}_{i})\rho_{sj}- (k^{\mu}F_{\mu\nu}\epsilon_{j}^{\nu}\  k^{\lambda}F_{\lambda\rho}\epsilon^{\ast\rho}_s)\rho_{is}].
\end{align}
In the following, we choose $\vec k=k_{0}\hat k$, where $\hat k$ indicates direction of outgoing photons from the transverse probe laser. We use the transverse condition for  real polarization vectors, $\vec \epsilon_{1}(k)$ and  $\vec \epsilon_{2}(k)$ being  orthogonal to the direction of $\vec k$.  The background antisymmetric field tensor $F_{\mu\nu}$  is expressed in terms of electric and magnetic field components
\begin{align}\label{ee25}
F_{\mu\nu}=\begin{pmatrix}0 & E_{x} & E_{y} & E_{z} \\
-E_{x} & 0 & B_{z} & -B_{y} \\
-E_{y}& -B_{z} & 0 & B_{_{x}} \\
-E_{z}& B_{y} & -B_{x} & 0 \\
\end{pmatrix}
\end{align}
The magnetic-field components can be represented as a vector field $\vec B=F_{23}\hat x +F_{31}\hat y+F_{12}\hat z$; then we can proceed to the calculation for example the term  $k^{i}F_{i j}\epsilon^{j}_{s} $ in Eq. (\ref{e25})
\begin{align}\label{eee25}
k^{i}F_{i j}\epsilon^{j}_{s}=F_{12}( k^{1} \epsilon^{2}_{s}-k^{2} \epsilon_{s}^{1})+F_{23}(k^{2} \epsilon_{s}^{3}-k^{3} \epsilon_{s}^{2})+F_{31}( k^{3}\epsilon^{1}_{s}-k^{1} \epsilon^{3}_{s})=\vec B\cdot( \vec k\times \hat \epsilon_{s}).
\end{align}
Also the electric-field components can be represented as a vector field $ \vec E=F_{01}\hat x +F_{02}\hat y+F_{03}\hat z$ and
\begin{align}\label{e25a}
k^{0}F_{0 i}\epsilon_{s}^{i}=k^{0}(F_{01}\epsilon_{s}^{1}+F_{02}\epsilon_{s}^{2}+F_{03}\epsilon_{s}^{3})
=k^{0}\vec E \cdot\hat \epsilon_{s}.
\end{align}
As a result the  time derivative components of  the density matrix can be expressed as
\begin{align}\label{e25c}
\frac{d\rho_{ij}}{dt}=\frac{2i\alpha^{2}k^{0}}{15m^{4}_{e}}\Big[\sum_{s} \Big(\vec B \cdot(\hat k\times\hat \epsilon_{s})\ \vec B\cdot(\hat k\times\hat\epsilon_{i^{}})+\vec E\cdot\hat \epsilon_{s}\vec B\cdot(\hat k\times\hat\epsilon_{i^{}})+\vec E\cdot\hat \epsilon _{i} \vec B\cdot(\hat k\times\hat\epsilon_{s^{}})+\vec E\cdot\hat \epsilon _{i} \vec E\cdot\hat \epsilon_{s}\Big)\rho_{sj}\\ \nonumber- \Big(\vec B\cdot(\hat k\times\hat\epsilon_{j})\ \vec B\cdot(\hat k\times\hat \epsilon_{s^{}})+\vec E\cdot\hat \epsilon_{j}\vec B\cdot(\hat k\times\hat\epsilon_{s^{}})+\vec E\cdot\hat \epsilon _{s} \vec B\cdot(\hat k\times\hat\epsilon_{j^{}})+\vec E\cdot\hat \epsilon _{j} \vec E\cdot\hat \epsilon_{s}\Big)\rho_{is}\Big]
\end{align}
the components of which are
\begin{align}\label{e26}
\dot \rho_{11 }^{}=\frac{2i\alpha^{2}k^{0}}{15m^{4}_{e}} \Big[\Big(\vec B \cdot(\hat k\times\hat \epsilon_{2})\ \vec B\cdot(\hat k\times\hat\epsilon_{1^{}})+\vec E\cdot\hat \epsilon_{2}\vec B\cdot(\hat k\times\hat\epsilon_{1^{}})+\vec E.\hat \epsilon _{1} \vec B\cdot(\hat k\times\hat\epsilon_{2^{}})+\vec E\cdot\hat \epsilon _{1} \vec E\cdot\hat \epsilon_{2}\Big) (\rho_{21}-\rho_{12})\Big]
\end{align}

\begin{align}\label{e27}
\dot \rho_{22 }^{}=\frac{2i\alpha^{2}k^{0}}{15m^{4}_{e}} \Big[\Big(\vec B \cdot(\hat k\times\hat \epsilon_{2})\ \vec B\cdot(\hat k\times\hat\epsilon_{1^{}})+\vec E\cdot\hat \epsilon_{2}\vec B\cdot(\hat k\times\hat\epsilon_{1^{}})+\vec E\cdot\hat \epsilon _{1} \vec B\cdot(\hat k\times\hat\epsilon_{2^{}})+\vec E\cdot\hat \epsilon _{1} \vec E\cdot\hat \epsilon_{2}\Big) (\rho_{12}-\rho_{21})\Big]
\end{align}
\begin{align}\label{e28}
\dot \rho_{12 }^{}=\frac{2i\alpha^{2}k^{0}}{15m^{4}_{e}} \Big[\Big(\vec B \cdot(\hat k\times\hat \epsilon_{2})\ \vec B\cdot(\hat k\times\hat\epsilon_{1^{}})+\vec E\cdot\hat \epsilon_{2}\vec B\cdot(\hat k\times\hat\epsilon_{1^{}})+\vec E\cdot\hat \epsilon _{1} \vec B\cdot(\hat k\times\hat\epsilon_{2^{}})+\vec E\cdot\hat \epsilon _{1} \vec E\cdot\hat \epsilon_{2}\Big) (\rho_{22}-\rho_{11})\\ \nonumber+\Big((\vec B\cdot\hat k\times\hat \epsilon_{1}+\vec E\cdot\hat \epsilon_{1})^{2}-(\vec B\cdot\hat k\times\hat \epsilon_{2}+\vec E\cdot\hat \epsilon_{2})^{2}\Big)\rho_{12}\Big]
\end{align}
\begin{align}\label{e29}
\dot \rho_{21 }^{}=\frac{2i\alpha^{2}k^{0}}{15m^{4}_{e}}\Big[\Big(\vec B \cdot(\hat k\times\hat \epsilon_{2})\ \vec B\cdot(\hat k\times\hat\epsilon_{1^{}})+\vec E\cdot\hat \epsilon_{2}\vec B\cdot(\hat k\times\hat\epsilon_{1^{}})+\vec E\cdot\hat \epsilon _{1} \vec B\cdot(\hat k\times\hat\epsilon_{2^{}})+\vec E\cdot\hat \epsilon _{1} \vec E\cdot\hat \epsilon_{2}\Big) (\rho_{11}-\rho_{22})\\ \nonumber+\Big((\vec B\cdot\hat k\times\hat \epsilon_{2}+\vec E\cdot\hat \epsilon_{2})^{2}-(\vec B\cdot\hat k\times\hat \epsilon_{1}+\vec E\cdot\hat \epsilon_{1})^{2}\Big)\rho_{21}\Big]
\end{align}
Hence  the time evolution of the Stokes parameters is:
\begin{align}\label{e30}
\dot I=\dot \rho_{11}+\dot \rho_{22}=0
\end{align}
\begin{align}\label{e31}
\dot Q=\dot \rho_{11}-\dot \rho_{22}=-\frac{4\alpha^{2}k^{0}}{15m^{4}_{e}} \Big[\Big(\vec B \cdot(\hat k\times\hat \epsilon_{2})\ \vec B\cdot(\hat k\times\hat\epsilon_{1^{}})+\vec E\cdot\hat \epsilon_{2}\vec B\cdot(\hat k\times\hat\epsilon_{1^{}})+\vec E\cdot\hat \epsilon _{1} \vec B\cdot(\hat k\times\hat\epsilon_{2^{}})+\vec E\cdot\hat \epsilon _{1} \vec E\cdot\hat \epsilon_{2}\Big)\Big]V
\end{align}
\begin{align}\label{e32}
\dot U=\dot \rho_{21}+\dot \rho_{12}=-\frac{2\alpha^{2}k^{0}}{15m^{4}_{e}} \Big[(\vec B\cdot\hat k\times\hat \epsilon_{2}+\vec E\cdot\hat \epsilon_{2})^{2}-(\vec B\cdot\hat k\times\hat \epsilon_{1}+\vec E\cdot\hat \epsilon_{1})^{2}\Big]V
\end{align}
\begin{align}\label{e33}
\dot V=i(\dot \rho_{12}-\dot \rho_{21})=\frac{2\alpha^{2}k^{0}}{15m^{4}_{e}} \Big[2\Big(\vec B \cdot(\hat k\times\hat \epsilon_{2})\ \vec B.(\hat k\times\hat\epsilon_{1^{}})+\vec E\cdot\hat \epsilon_{2}\vec B\cdot(\hat k\times\hat\epsilon_{1^{}})+\vec E\cdot\hat \epsilon _{1} \vec B\cdot(\hat k\times\hat\epsilon_{2^{}})+\vec E\cdot\hat \epsilon _{1} \vec E\cdot\hat \epsilon_{2}\Big)Q\\ \nonumber +\Big((\vec B\cdot\hat k\times\hat \epsilon_{2}+\vec E\cdot\hat \epsilon_{2})^{2}-(\vec B\cdot\hat k\times\hat \epsilon_{1}+\vec E\cdot\hat \epsilon_{1})^{2}\Big)\Big]U
\end{align}
Equation (\ref{e30}) tells us that, in the photons ensemble, the total intensity of photons does not depend on the photon-photon forward-scattering term. According to Eqs.
(\ref{e31})-(\ref{e33}), the unpolarized photons (namely, $Q=U=V=0$) can not acquire any polarization during  propagating through the background fields. In contrast  linear polarized photons (namely, \emph{Q} and/or \emph{U} $\neq$0 )  can  acquire circular polarization $V\neq0$ [see Eq. (\ref{e33})]. We recall that the generation of circular polarizations from linearly polarized laser beam collisions due to the Euler-Heisenberg effective Lagrangian was discussed in  \cite{2014PhRvA..89f2111M}.

\section{Analytical  And Numerical Solutions}

In the last section we investigated  the generation of  circular polarization in the presence of strong background fields. Here we attempt to consider two cases. The strong background field is  made by (i) an ultra-intense laser pulse(time dependent fields), and (ii)  superconductor magnets (static fields). The time evolution of Stokes parameters   (\ref{e30})-(\ref{e33}) can be rewritten as below

\begin{align}\label{e34}
\dot I=0, \ \ \  \  \dot Q=-\Omega_{QV}V, \ \ \ \  \dot U=-\Omega_{UV}V, \ \  \  \ \dot V=\Omega_{QV}Q+\Omega_{UV}U,
\end{align}
\begin{align}\label{e35}
\Omega_{QV}=\frac{4\alpha^{2}k^{0}}{15m^{4}_{e}} \Big[\Big(\vec B \cdot(\hat k\times\hat \epsilon_{2})\ \vec B\cdot(\hat k\times\hat\epsilon_{1^{}})+\vec E\cdot\hat \epsilon_{2}\vec B\cdot(\hat k\times\hat\epsilon_{1^{}})+\vec E\cdot\hat \epsilon _{1} \vec B\cdot(\hat k\times\hat\epsilon_{2^{}})+\vec E\cdot\hat \epsilon _{1} \vec E\cdot\hat \epsilon_{2}\Big)\Big]
\end{align}
\begin{align}\label{e36a}
\Omega_{UV}= \frac{2\alpha^{2}k^{0}}{15m^{4}_{e}} \Big[(\vec B.\hat k\times\hat \epsilon_{2}+\vec E.\hat \epsilon_{2})^{2}-(\vec B.\hat k\times\hat \epsilon_{1}+\vec E.\hat \epsilon_{1})^{2}\Big]
\end{align}
From these equations we have 
\begin{align}\label{e37}
\ddot V=-\Omega ^{2}V+\dot \Omega _{QV} Q+\dot \Omega _{UV} U,
\end{align}
\begin{align}\label{e38}
\Omega ^{2}=\Omega^{2} _{QV} +\Omega^{2} _{UV}.
\end{align}
From Eq. (\ref{e34}) it is easy to show $Q\dot Q+U\dot U+V\dot V=0$, and the  relation between Stokes parameters  ($I^{2}=Q^{2}+U^{2}+V^{2}$)  is automatically satisfied.
\\ 
\subsection{ Analytical solution in the presence of time-independent background field}\label{ANA}

In the case of time-independent background fields,  the $\Omega_{QV}$ and $\Omega_{UV}$ are constant in time, and the Eq. (\ref{e37}) becomes  a simple harmonic equation $\ddot V+\Omega ^{2}V=0$ which has a general solution
\begin{align}\label{a39}
V(t)=\mathcal{A}\sin(\Omega t)+\mathcal{B}\cos(\Omega t).
\end{align}
The coefficients $\mathcal{A}$ and $\mathcal{B}$ are determined by initial conditions at $t=0$ 

\begin{align}\label{a40}
V(0)=\mathcal{B}\ \ \ \ \ \ \ \  \dot V(0)=\mathcal{A}\Omega=\Omega_{QV}Q(0)+\Omega_{UV}U(0).
\end{align}
It is clear from Eq. (\ref{a40}) that $\dot V(0)$ is determined by linear polarization parameters $Q(0)$ and $U(0)$. $Q(t)$ and $U(t)$ can be found by adopting  Eqs. $(\ref{e34})$ and $(\ref{a39})$ as follows
\begin{align}\label{a43}
\dot Q=-\Omega_{QV}V\Longrightarrow Q(t)=\frac{\Omega_{QV}}{\Omega}[\mathcal{A}\cos(\Omega t)-\mathcal{B}\sin(\Omega t)],
\end{align}
\begin{align}\label{a44a}
\dot U=-\Omega_{UV}V\Longrightarrow U(t)=\frac{\Omega_{UV}}{\Omega}[\mathcal{A}\cos(\Omega t)-\mathcal{B}\sin(\Omega t)],
\end{align}
we supposed a totally linear polarized incoming radiation with $P_{0}=1$ $(Q_{0}=U_{0}={1}/{\sqrt{2}}$) without any initial circular polarization ($V_{0}=0$), then the time evolution of Stokes parameters is given by
\begin{align}\label{a44}
V(t)=\sin(\Omega t), \ \ \ U(t)=Q(t)=\frac{1}{\sqrt{2}}\cos(\Omega t).
\end{align}
The final results in the time-independent background field show that  the Stokes parameters are harmonically oscillating in time. This means that we should not expect any circular component after long time with respect to the period  $\Omega^{-1}$. However, one could measure the  time average of squared total linear polarization $\left\langle P^{2} \right\rangle=\left\langle Q^{2} \right\rangle+\left\langle U^{2} \right\rangle =1$ after several periods.  
The frequency $\Omega$ of these oscillations can be estimated in a real experimental setup like PVLAS(polarizzazione del vuoto con laser) experiments \cite{2008PhRvD..77c2006Z,2013NJPh...15e3026D,2016EPJC...76...24D},  where a linear polarized laser  beam with wavelength of 1064 nm  propagates through  a static magnetic field of about $2. 5 \ \mathrm{T}$, the direction is orthogonal to the laser beam direction. With the help of Eqs. (\ref{e35}),(\ref{e36a}) and (\ref{e38}) we have
\begin{align}\label{a45}
\Omega =\frac{2\alpha^{2}k^{0}B^2_{0}}{15m^{4}_{e}}=\frac{\alpha}{15\lambda_{0}}(\frac{eB_{0}}{m^{2}_{e}})^{2}=4.43\times 10^{-8}s^{-1}.
\end{align}
Then the period $\mathrm{T}=1.12\times 10^{8}s$ for one period or corresponding to the length $\mathrm{L}=3.36\times 10^{16}m$. However the magnets of the PVLAS experiment  have a total magnetic field length $\mathrm{L}=1.6m$, where $\mathrm{L}$ is the optical path length within the birefringent region. This means that  the oscillating effect [Eq. (\ref{a44})] does not appear in the PVLAS like  experiments. Let
us determine the ellipticity signal  by using Eqs. (\ref{e017}) and (\ref{a44})
\begin{align}\label{a46}
\epsilon_{QED}=\frac{1}{2}\tan ^{-1}\Big(\frac{\sin(\Omega t)}{\cos(\Omega t)}\Big)=\frac{\Omega t}{2}=\frac{\alpha}{30}(\frac{eB}{m^{2}_{e}})^{2}\frac{L}{\lambda_{0}}=1.18\times10^{-16}.
\end{align}
This value of the ellipticity is very far from our current detectors precision. We note that the optical path length can be increased by using a Fabry-Perot cavity  of finesse $\mathcal{F}$ as discussed in \cite{2013NJPh...15e3026D}, where one can define  the effective path length $\mathrm{L}_{eff}=\frac{2\mathcal{F}L}{\pi}$. With finesses $\mathcal{F}>400000$  the value of  the ellipticity signal can be increased up to  $10^{-11}$.
We use  Eqs. (\ref{e016}) and (\ref{a44}) to obtain the rotation of the polarization plane 
\begin{align}\label{a146}
\psi_{QED}=\frac{1}{2}\tan ^{-1}\Big(\frac{U}{Q}\Big)= \frac{1}{2}\tan ^{-1}\Big(1\Big) \ \  \Longrightarrow  \  \ \dot \psi_{QED}=0.
\end{align}
These results confirm  previous studies \cite{1971AnPhy..67..599A,1998ftqp.conf...29D,2006OptCo.267..318H,1997JPhA...30.6485H} which are mainly based on  the semi-classical approach (viewing fields as classical fields) and evaluation of one loop polarization tensor.  Our approach is based on using the quantum Boltzmann equation to determine all the  Stokes parameters regarding the quantum structure of a probe beam. In contrast to previous consideration where the assumption of the constant background field  has been used to extract the photon polarization tensor from the Euler-Heisenberg Lagrangian, our method can be easily extended to time-dependent background fields ( see Sec. \ref{NUM}).
\\ It is clear from Eq. (\ref{a146}) that the nonlinear QED interactions through the Euler-Heisenberg Lagrangian can not rotate the linear polarization plane for a laser beam traversing a static magnetic field. Regarding Eq. (\ref{e010}) the rotation angle $\psi$ of an initially linear polarized photon is due to the imaginary part of the refractive index ($\tilde n=n+i\kappa$). The effects which modify the light amplitudes in the different directions will induce a rotation angle. Those effects can not be explained in QED below the threshold ($\omega< 2m_{e}$) where the electron-positron pair production rate is exponentially suppressed and further possibilities such as photon splitting \cite{1971AnPhy..67..599A} or neutrino-pair production \cite{2000PhLB..480..129G} are unmeasurably small. In other words QED does not predict dichroism. Therefore, a non-negligible signal for  vacuum rotation angle $\psi$ may imply evidence for new physics beyond the standard model of particle physics. This  could be interpreted as a neutral scalar or pseudo-scalar particle weakly coupling to two photons called axion-like particles (ALPs) \cite{1986PhLB..175..359M,1987PhRvL..59..396G} or other  candidates such as millicharged particles (MCPs) \cite{2006PhRvL..97n0402G,2007PhRvD..75c5011A}. It is worth mentioning that  an observation of the rotation of  polarization plane by the vacuum in a static magnetic field was reported by PVLAS group in 2006 \cite{2006PhRvL..96k0406Z}; they announced the existence of a very light, neutral, spin-zero particle coupled to two photons, while  this observation was excluded by data taken with an upgraded setup in 2008 \cite{2008PhRvD..77c2006Z}.
 \subsection{Numerical solution in the case of  two laser beams collisions}\label{NUM}

\begin{figure}
    \centering
	\vspace{-2cm}
    {
        \includegraphics[width=0.47\textwidth]{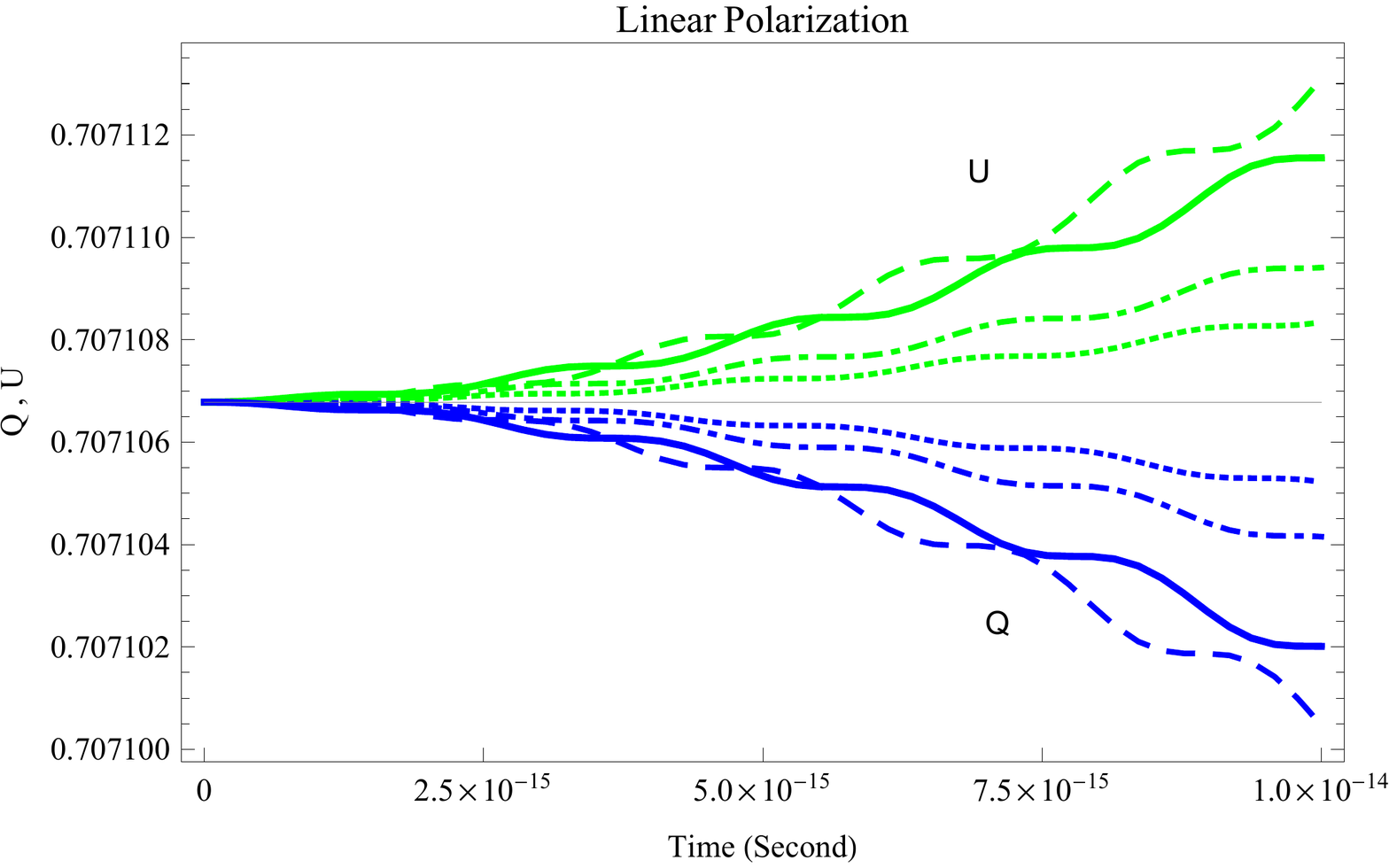}
}
 {  \includegraphics[width=0.47\textwidth]{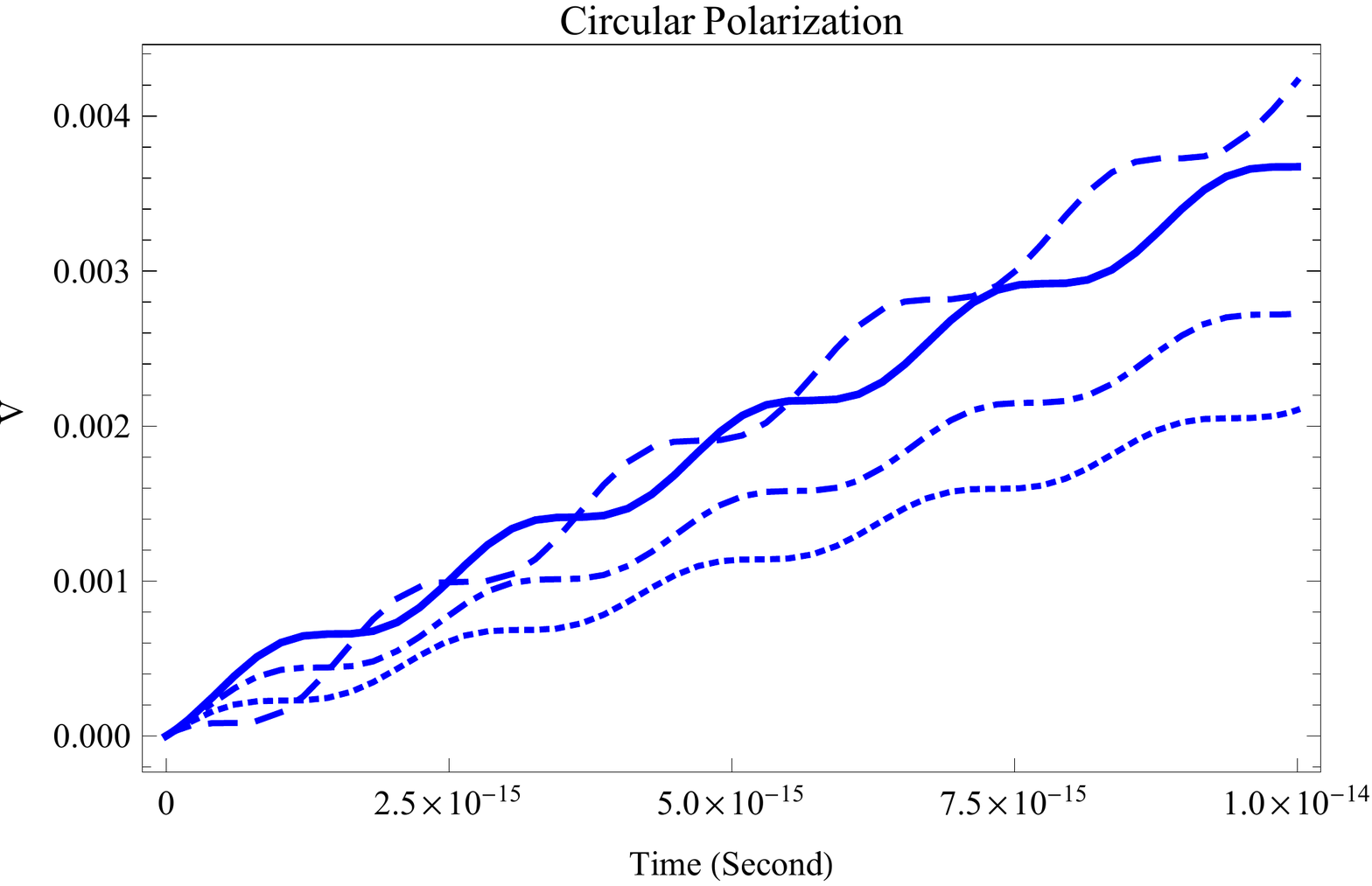} }
 \vspace{-2.8cm}
\caption{\scriptsize (Color online) In the left panel,  the time evolution of dimensionless  Stokes parameters \emph{U} [the upper half plane (green )] and \emph{Q} [the lower half plane (blue)].  In the right panel,  the time evolution of dimensionless Stokes parameter \emph{V}. They are plotted for a 10 keV linearly polarized probe laser beam interacting with a target laser beam in optical frequency $\omega=1eV$ and peak intensity  $I=3\times10^{22}{W}/{cm^{2}}$. These figures show  the impact of collision geometry for two laser beams on the evolution of Stokes parameters, for different polar angles $\theta=\pi$ (dashed lines), ${\pi}/{2}$ (dotted lines), ${\pi}/{3}$ (dot-dashed lines), and ${\pi}/{5}$ (solid lines) at a fixed azimuthal angle  $
\phi={\pi}/{8}$.}\label{Fig3}
\end{figure}

\begin{figure}
    \centering
	\vspace{-2.5cm}
    {
        \includegraphics[width=0.47\textwidth]{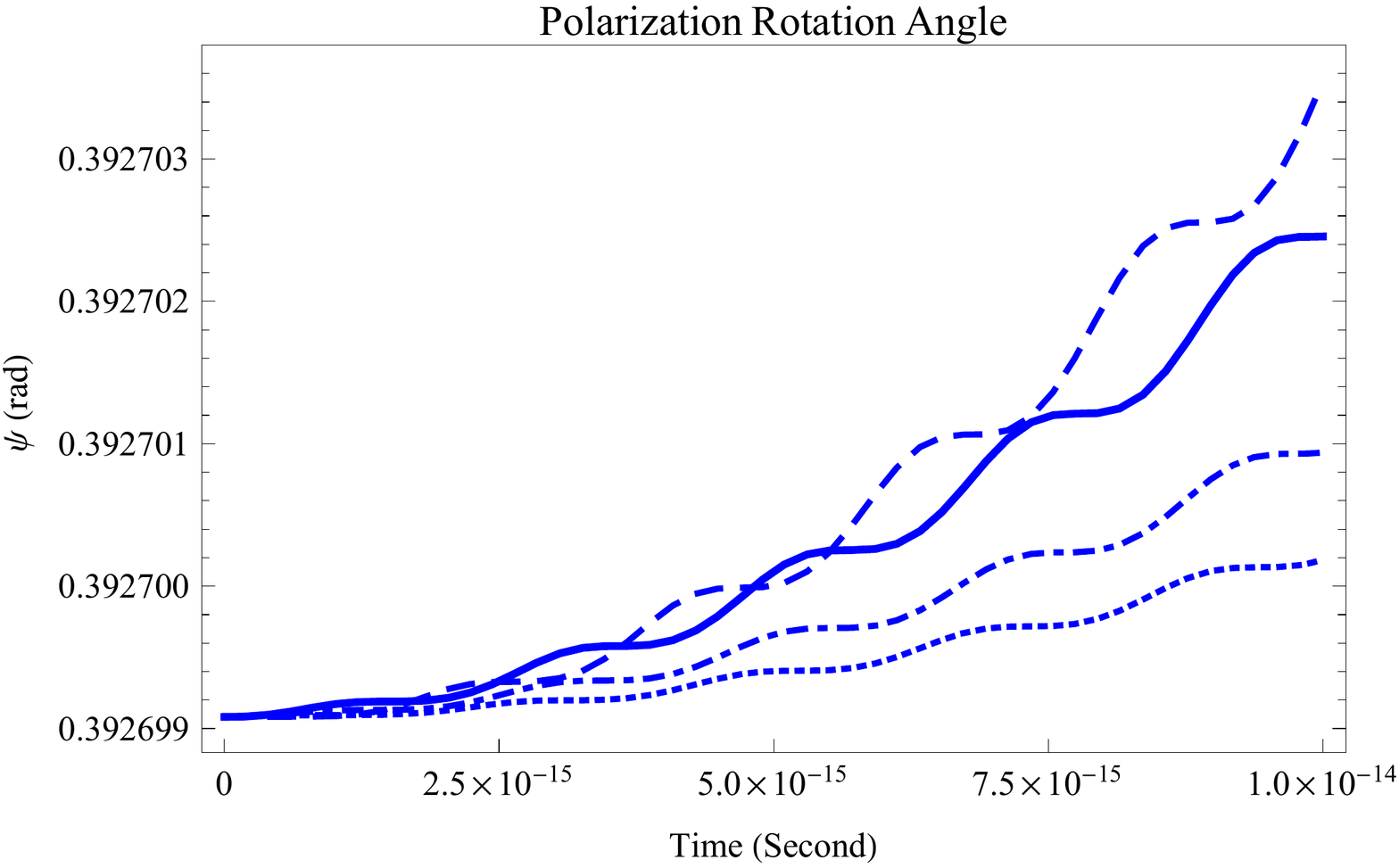}
}
 {  \includegraphics[width=0.47\textwidth]{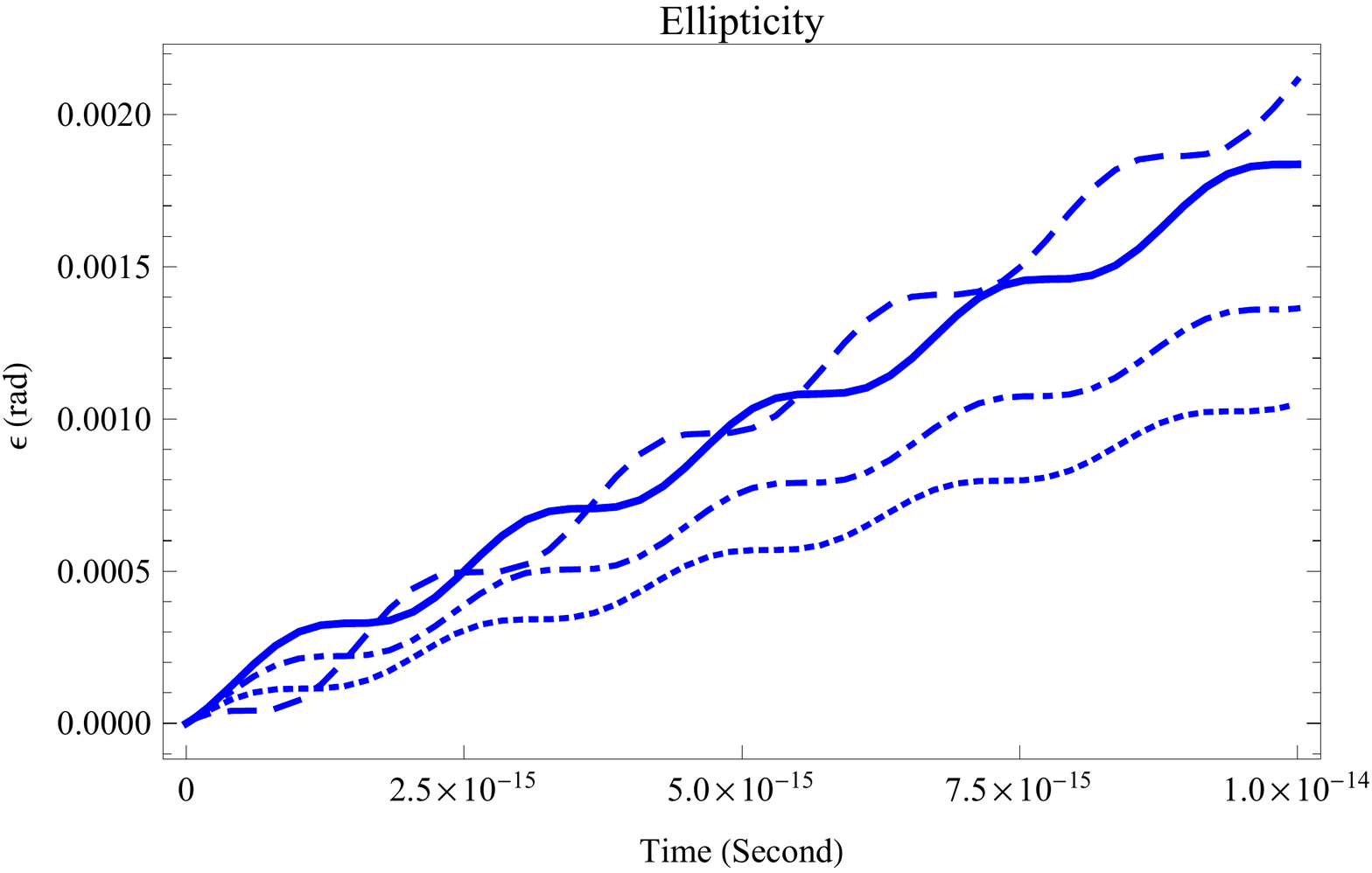} }
\vspace{-2.8cm}
\caption{\scriptsize (Color online) In the left panel, the polarization rotation angle $\Psi$ and in the right panel, the ellipticity parameter $\epsilon$ as a function of time for the cases depicted in Fig. \ref{Fig3}}\label{Fig4}
\end{figure}

In  this section, we want to consider collision of two laser beams. Assuming that the linearly polarized probe beam  passes through another laser beam with lower frequency,  the latter is considered as a background field. Consider the collision of two linearly polarized laser pulses, a low-intensity x-ray probe pulse $(\hat k,\hat \epsilon_{1}(k),\hat \epsilon_{2}(k))$ crossing the high-intensity optical pulse $(\hat p,\hat E(p),\hat B (p))$ as a target beam. Due to the separation in energy scales the probe beam essentially scatters forward. Here in contrast to the previous section the background fields vary with time. The  electric and magnetic fields  are oscillating, their amplitudes are related to each other by $\left\vert E \right\vert=\left\vert B \right\vert$, and directions are orthogonal to the beam direction
\begin{align}\label{e322}
 \ \  \hat k=\begin{pmatrix} 0 \\
 0 \\
1 \\
\end{pmatrix}\  \ \hat \epsilon_{1}(k)^{}=\begin{pmatrix} 1 \\
 0 \\
0\\
\end{pmatrix} \ \ \hat \epsilon_{2}(k)^{}=\begin{pmatrix} 0 \\
 1 \\
0 \\
\end{pmatrix},
\end{align}
and
\begin{align}\label{e323}
 \ \  \hat p=\begin{pmatrix} \sin \theta \cos \phi\\
 \sin \theta \sin \phi \\
\cos \theta \\
\end{pmatrix}\  \ \hat E(p)^{}=\sin \omega t\begin{pmatrix} \cos \theta \cos \phi \\
 \cos \theta \sin \phi\\
-\sin \theta \\
\end{pmatrix} \ \ \hat B(p)^{}=\cos \omega t\begin{pmatrix} -\sin \phi \\
 \cos \phi\\
0 \\
\end{pmatrix},
\end{align}
where $\hat k$, $\hat \epsilon_{1}$ and $\hat \epsilon_{2}$ are momentum  and polarization unit vectors of the incident (probe) beam; $\hat p$, $\hat E$ and $\hat B$ are momentum, electric and magnetic field unit vectors of the target laser beam ($\vec E=E_{0}\hat E,\vec B=B_{0}\hat B$ ). Regarding  Eqs. (\ref{e322}) and (\ref{e323}) for $\Omega _{QV} $ and $\Omega _{UV} $ we  obtain:
\begin{align}\label{e35a}
\Omega_{QV}=G\sin 2\phi[\cos^{2}\omega t+\sin2\omega t\cos\theta+\sin^{2}\omega t\cos^{2}\theta],
\end{align}
\begin{align}\label{e36}
\Omega_{UV}=-G\cos 2\phi[\cos^{2}\omega t+\sin2\omega t\cos\theta+\sin^{2}\omega t\cos^{2}\theta],
\end{align}
Where
\begin{align}\label{ea37}
G=[\frac{2\alpha^{2}k^{0}B^2_{0}}{15m^{4}_{e}}]=\frac{2.99\alpha}{15}[\frac{\mathrm{I}}{\mathrm{I}_{c}}][\frac{1 nm}{\lambda_{0}}]\times10^{20}s^{-1}=1.45 [\frac{\mathrm{I}}{\frac{W}{cm^{2}}}][\frac{1nm}{\lambda_{0}}]\times10^{-12}s^{-1}.
\end{align}
We assume that the  background field  is an intense and focused laser beam in optical frequency  $\omega=1$ $eV$,  and this optical petawatt laser has  the peak intensity $I={P}/{\pi d^{2}}\simeq3\times10^{22}{W}/{cm^{2}}$ (power \emph{P}=\emph{1 PW} and  pulse length d=$1\mu m$). It should be noted that  $(eB)^{2}=({\mathrm{I}}/{\mathrm{I}_{c}})(eB)^{2}_{c}$ where ${I}_{c}=10^{29}{W}/{cm^{2}}$ and $(eB)_{c}=m^{2}_{e}=1.282\times 10^{13}Gauss$. Let  us assume the  probe beam is an  XFEL,  with $\lambda$=0.1 nm wave length, hence $k^{0}=h\nu ^{0}=10 keV$, like the DESY project \cite{HIBEF}. At x-ray photon energies of the order of 10 keV, the
highest polarization purities can be achieved by reflection
at perfect crystals at a Bragg angle of $\pi/4$ \cite{Marx:2013db}. This x-ray probe beam can also be achieved experimentally from a laser-based Thomson back-scattering source \cite{2006OptCo.267..318H}. It has been supposed that the initial value of the  circular polarization $V_{0}=0$ and the linear polarization $P_{0}=\sqrt{Q_{0}^{2}+U_{0}^{2}}=1$ ($Q_{0}=U_{0}={1}/{\sqrt{2}}$), which are all dimensionless quantities. The Stokes parameters are normalized by the intensity \emph{I}, giving the dimensionless quantities. Using this set of  parameters we solved Eqs. (\ref{e34}) and determined the time evolution of  the Stokes parameters followed by ellipticity and polarization rotation angles. The results for various polar angles ($\theta$) of   two beams direction in a fixed  azimuthal angle  $\phi={\pi}/{8}$ are displayed in Fig. (\ref{Fig3}) and (\ref{Fig4}). These figures show the polarization evolution of an X-ray probe beam as result of interacting  with the optical laser field in  the crossing region of two laser beams. The interaction time is limited by the pulse length d=$1\mu m$. As one can find from Fig. (\ref{Fig3}) (right panel), the circular polarization \emph{V} gets its maximum value for the head on collision of the two laser beams ($\theta=\pi$), and its minimum value for  transverse collision angle ($\theta=\pi/2$). From Fig.  (\ref{Fig3}) (left panel), it is obvious that U increases with time although \emph{Q} decreases with time, according to Eqs. (\ref{e34}), the change of \emph{Q} and \emph{U} with time acts as a source term for circular polarization parameter \emph{V}.

In the presence of time-dependent background field, the governing equation of Stokes parameter \emph{V} is not a simple harmonic equation [ see Eq. (\ref{e37})], since $\Omega_{QV}$ and $\Omega_{UV}$ of Eqs. (\ref{e35a}) and (\ref{e36})  are time-dependent quantities in this case. In the laser-laser collision  ${U}/{Q}$ is a function of time as a result of time-dependent configuration in the interaction region which leads to a time-evolving rotation angle $\psi_{QED}(t)$
\begin{align}\label{e307}
\frac{\dot U}{\dot Q}=\frac{\Omega_{UV}(t)}{\Omega_{QV}(t)} \ \  \Longrightarrow  \  \  \dot \psi_{QED}= \frac{d}{dt}\Big(\frac{1}{2}\tan ^{-1}\Big(\frac{U}{Q}\Big)\Big)\neq 0.
\end{align}
This result is different from what we obtained in the propagation of a laser beam in the static magnetic field of PVLAS-like experiments [ see Eq. (\ref{a146})]. The rotation angle of polarization plane acquired by  a probe beam via crossing region of size $1\mu m$ is $\Delta\psi\sim4.5\times10^{-6}$ rad in the head on collision ($\theta=\pi$), as shown  in  Fig.  (\ref{Fig4}) (left panel). 
This result is  considered here in QED below  the threshold energy of pair production. Our approach provides  an explanation for vacuum  rotation within the framework of QED in the  presence of time-dependent background field. Therefore, it shows us that any sizable signal of the vacuum rotation angle $\psi$ in the strong-field regime is not essentially a signature of new physics (ALP's and MCP's). Besides, Fig.  (\ref{Fig4}) (right panel) shows that the maximum value of the ellipticity $\epsilon$ acquired by probe photons in the above-mentioned interaction length is $2\times10^{-3}$ rad, which is three orders of magnitude larger than the rotation angle. Moreover the light diffraction by a strong standing electromagnetic wave also leads to vacuum rotation but for pure kinematical reasons, depending on the details of the optical setup \cite{2009EPJD...55..311G,2006PhRvL..97h3603D}. This is different from the laser-laser collisions which we are considering here. The rotation angle found in \cite{2006PhRvL..97h3603D} is the same order of magnitude as the induced ellipticity.
Thanks to recent technological advances in x-ray polarimetry \cite{Marx:2013db}, the x-ray polarimeters allow detection of rotation of the polarization plane down to 1 arcsec ($4.8\times10^{-6}$ rad) and ellipticities of about $1.51\times10^{-5}$ rad. Recently, authors in \cite{2016PhyS...91b3010S} presented a comprehensive study of  the feasibility of measuring vacuum birefringence by probing the focus of a high intensity optical laser with an x-ray free-electron laser. 
\\ In this paper our computations are based on perfect vacuum assumption, actually  the presence of charged particles in the interaction region may obscure the measurement of real photon-photon scattering \cite{2015arXiv151008456K}. Since in the presence of an electromagnetic wave, a gas becomes a birefringent medium due to the Cotton-Mouton and Kerr effects \cite{2008PhRvD..77c2006Z,Rizzo:2010jd}, it is compulsory to clean the interaction region by removing all residual gas particles to avoid an additional background signal. Fortunately, because of the small interaction region in the laser-laser collision, the absence of residual particles is only required for a short time period \cite{2016PhyS...91b3010S}.

\section{Polarization of a probe laser beam in time-dependent and static background fields}

\begin{figure}
    \centering
	\vspace{-2.5cm}
    {
        \includegraphics[width=0.47\textwidth]{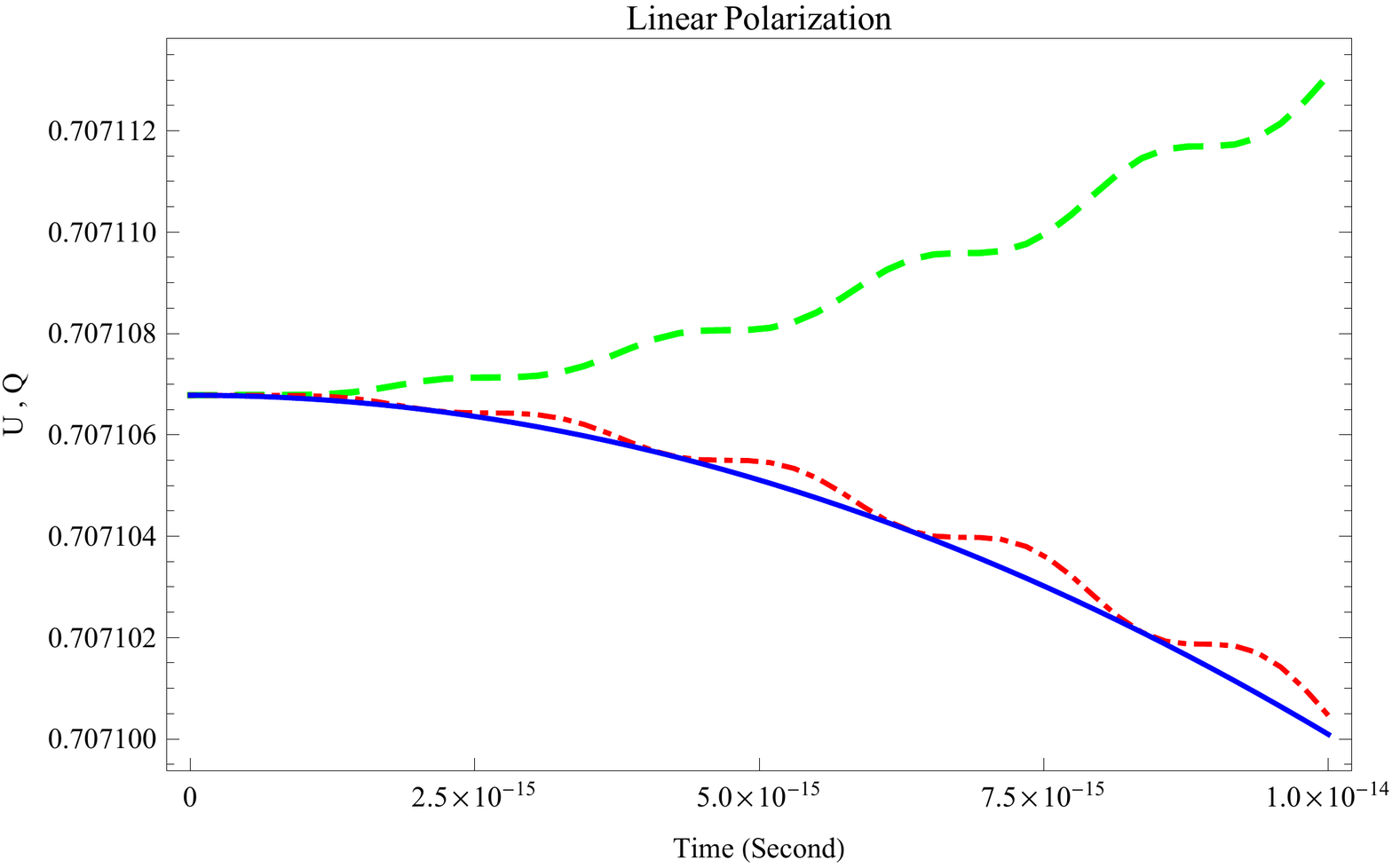}
}
 {  \includegraphics[width=0.47\textwidth]{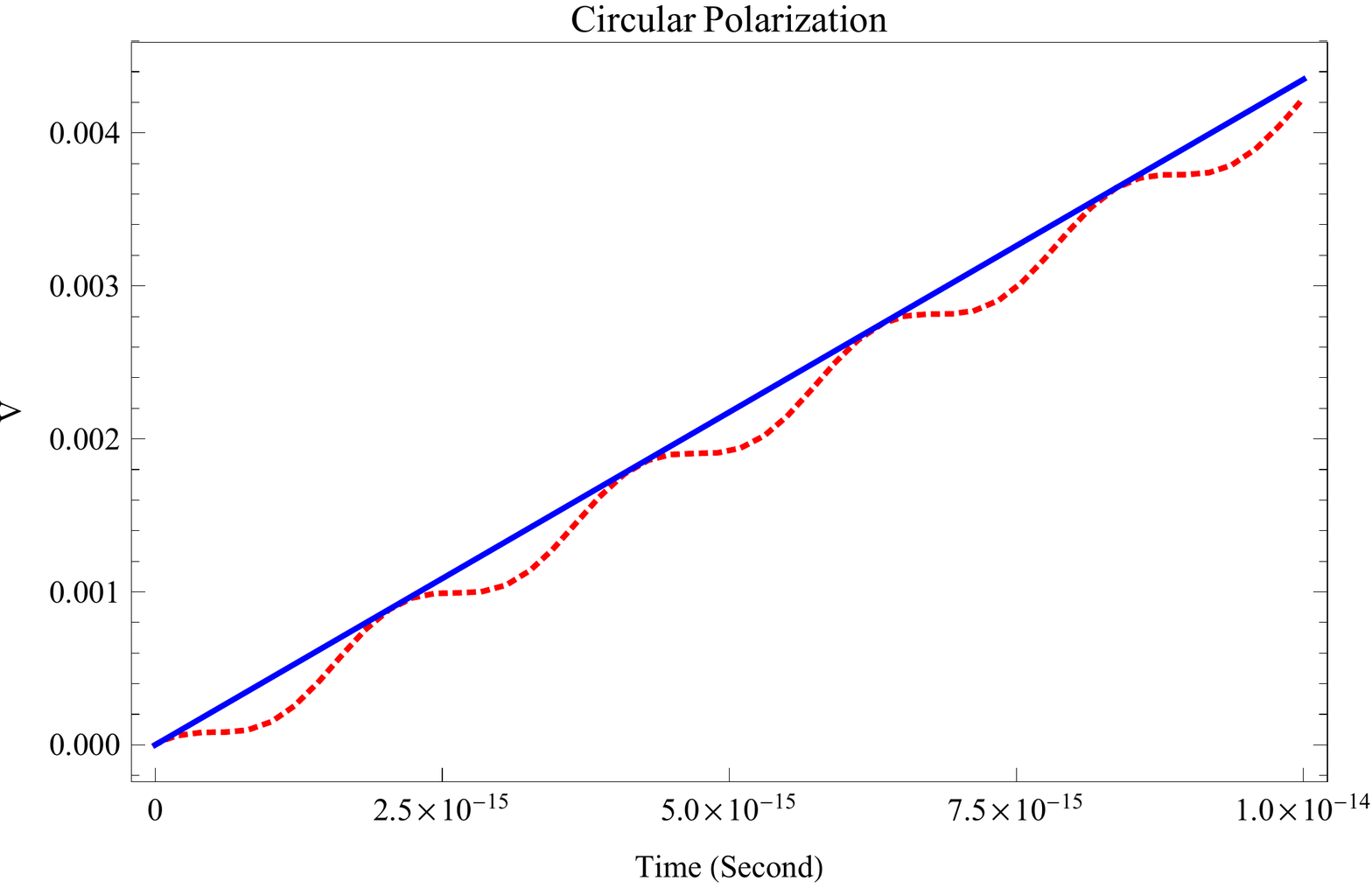} }
\vspace{-2.8cm}
\caption{\scriptsize  (Color online) Comparison between dimensionless Stokes parameters \emph{U}, \emph{Q} and \emph{V} in both time-dependent and static background fields. In time dependent case we used numerical solution of Sec. \ref{NUM} to plot \emph{U} [dashed (green) line] and \emph{Q} [dot-dashed (red) line] in the left panel and for \emph{V} [dotted (red) line] in the right panel. In the static magnetic field we have used analytic solution of Sec. \ref{ANA} to plot \emph{Q} and \emph{U} [solid (blue) line] in the left panel and \emph{V} [solid (blue) line] in the right panel. These figures are plotted for a 10 keV linearly polarized probe laser beam interacting with a target laser beam in optical frequency $\omega=1eV$ and peak intensity  $I=3\times10^{22}{W}/{cm^{2}}$.}\label{Fig5}
\end{figure}

In this section, we attempt to clarify the differences between using a static magnetic field  and  time-dependent fields that  induce polarization characteristics of a given probe laser beam. We assume a 10 keV probe laser beam with initial Stokes parameters $Q_{0}=U_{0}={1}/{\sqrt{2}}$ and $V_{0}=0$, and an optical target laser  beam with intensity $I=3\times10^{22}{W}/{cm^{2}}$ and frequency $\omega=1$ $eV$.  For the time-dependent fields which are realized in the cross section of laser beams, we used Eqs. (\ref{e34}) and (\ref{e322})-(\ref{e36}) with $\theta=\pi$ (head on collision) and $\phi=\frac{\pi}{8}$,  in which both  electric and magnetic fields are perpendicular to the beam direction $\hat k$ ($\hat E,\hat B\perp \hat k$). For the static magnetic field $B=7.02\times10^{9}Gauss $ corresponding to the  peak  intensity of  background optical laser $I=3\times10^{22}{W}/{cm^{2}}$, we used Eq. (\ref{a44}). In Fig. (\ref{Fig5}) we compare the evolution of Stokes parameters of the probe laser beam in these two cases (time-dependent and static background fields).  Here, we plotted analytic solutions  and numerical solutions  of Boltzmann equation for the same field intensity. Since  interaction time is too short ($10^{-14}$ s), we can approximate Eqs. (\ref{a44}) up to first order in time as
\begin{align}\label{a70}
V(t)=\sin(\Omega t)\approx \Omega t, \ \ \ U(t)=Q(t)=\frac{1}{\sqrt{2}}\cos(\Omega t) \approx \frac{1}{\sqrt{2}}.
\end{align}
This approximation shows us that \emph{V}  varies faster than \emph{Q} and/or \emph{U}  in a very short period of time. It helps us to explain three order-of-magnitude differences between ellipticity $\epsilon$ and polarization angle $\psi$  in the numerical solutions of laser-laser collision (Sec. \ref{NUM}). As  shown in Fig. (\ref{Fig5}) (left panel), in the static field there is not any conversion between \emph{U} and \emph{Q}, and then polarization angle will not change during propagation in such a static field, and the value of the \emph{V} parameter (right panel) in the presence of   time-dependent background field  oscillates around the value in  the static case.

\section{Conclusions and Remarks}
Due to the extremely large value of the critical field it remains very challenging to experimentally verify QED nonlinearities, by exploring polarization characteristics of the laser  photons propagation in the static and uniform magnetic field. As mentioned in this paper, much stronger electromagnetic fields can be produced by means of focused high power lasers. We have discussed the polarization properties induced in a probe laser beam during its propagation through a constant magnetic field or in collision with another laser beam.  We solved the quantum Boltzmann Equation  within the framework of Euler-Heisenberg Lagrangian for both time-dependent and static background fields to explore the time evolution of Stokes parameters \emph{Q}, \emph{U} and \emph{V}, [see  Eqs. (\ref{e34})]. It is shown that the oscillating solution for the Stokes parameters, in the static magnetic field has an oscillating period proportional to $k^{-1}_{0}B^{-2}_{0}$. The ellipticity signal can be increased up to $10^{-11}$ rad by using the cavity in the PVLAS experiment. Since it seems cavity techniques are inefficient for short pulsed high intensity fields to produce  standing wave like what is proposed in \cite{2006OptCo.267..318H,2006PhRvL..97h3603D}, it is necessary to consider  time-dependent  background fields. For the time-dependent background field we considered laser-laser collisions. We have generalized the conditions of previous work \cite{2014PhRvA..89f2111M} to the more realistic case of a  temporal laser background probed by x-ray photon pulses. We obtained maximum ellipticity $\epsilon\sim2\times10^{-3}$ rad and rotation of the polarization plane $\Delta\psi\sim4.5\times10^{-6}$ rad, in the case of high intensity background field $I=3\times10^{22}{W}/{cm^{2}}$ probed by 10 keV photon pulses. These values  are at the limit of the accuracy that can now be obtained with high-contrast x-ray polarimeters using multiple Bragg reflections from channel-cut perfect crystals \cite{2016PhyS...91b3010S,Alp:2000fr,Marx:2013db}. High-precision experiments using ultra-intense lasers would be testing the most successful theory QED in the strong field regime which has been little explored so far. Beyond this, these experiments can shed light on the hidden sector of the standard model, and such a test will be sensitive to physics beyond the standard model \cite{2006PhRvL..97n0402G,2008JPhA...41p4039G,PhysRevD.47.3707}.

\section{ACKNOWLEDGEMENTS}
S. Shakeri is grateful to Prof. R. Ruffini for supporting his visit at ICRANet Pescara, where the last part of this work is done. He would like to thank R. Mohammadi, M. Haghighat and C. Stahl for useful comments and discussions.

\end{document}